# Prospects for Multi-omics in the Microbial Ecology of Water Engineering

Elizabeth A. McDaniel[1*], Sebastian Aljoscha Wahl[2], Shun'ichi Ishii[3], Ameet Pinto[4], Ryan Ziels[5], Per H Nielsen[6], Katherine D. McMahon[1,7], Rohan B.H. Williams[8*]

[1] Department of Bacteriology, University of Wisconsin – Madison, Madison, WI, USA
[2] Department of Biotechnology, Delft University of Technology, Delft, The Netherlands
[3] Super-cutting-edge Grand and Advanced Research (SUGAR) Program, Institute for Extra-cutting-edge Science and Technology Avant-garde Research (X-star), Japan Agency for Marine-Earth Science and Technology (JAMSTEC), Yokosuka 237-0061, Japan
[4] Department of Civil and Environmental Engineering, Northeastern University, Boston, MA, USA
[5] Department of Civil Engineering, The University of British Columbia, Vancouver, BC, Canada
[6] Center for Microbial Communities, Aalborg University, Aalborg, Denmark
[7] Department of Civil and Environmental Engineering, University of Wisconsin – Madison, Madison, WI, USA
[8] Singapore Centre for Environmental Life Sciences Engineering, National University of Singapore, Republic of Singapore
* Corresponding authors: elizabethmcd93@gmail.com (E.A.M) and lsirbhw@nus.edu.sg (R.B.H.W)

**ABSTRACT**
Advances in high-throughput sequencing technologies and bioinformatics approaches over almost the last three decades have substantially increased our ability to explore microorganisms and their functions – including those that have yet to be cultivated in pure isolation. Genome-resolved metagenomic approaches have enabled linking powerful functional predictions to specific taxonomical groups with increasing fidelity. Additionally, related developments in both whole community gene expression surveys and metabolite profiling have permitted for direct surveys of community-scale functions in specific environmental settings. These advances have allowed for a shift in microbiome science away from descriptive studies and towards mechanistic and predictive frameworks for designing and harnessing microbial communities for desired beneficial outcomes. Water engineers, microbiologists, and microbial ecologists studying activated sludge, anaerobic digestion, and drinking water distribution systems have applied various (meta)'omics techniques for connecting microbial community dynamics and physiologies to overall process parameters and system performance. However, the rapid pace at which new 'omics-based approaches are developed can appear daunting to those looking to apply these state-of-the-art practices for the first time. Here, we review how modern genome-resolved metagenomic approaches have been applied to a variety of water engineering applications from lab-scale bioreactors to full-scale systems. We describe integrated 'omics analysis across engineered water systems and the foundations for pairing these insights with modeling approaches. Lastly, we summarize emerging 'omics-based technologies that we believe will be powerful tools for water engineering applications. Overall, we provide a framework for microbial ecologists specializing in water engineering to apply cutting-edge 'omics approaches to their research questions to achieve novel functional insights. Successful adoption of predictive frameworks in engineered water systems could enable more economically and environmentally sustainable bioprocesses as demand for water and energy resources increases.

**INTRODUCTION**
Microorganisms and complex microbial communities are an integral and essential component of engineered water systems designed to improve public health, maintain ecosystem health, and provide optimal water infrastructure and quality. Activated sludge (AS), anaerobic digestion (AD), and drinking water systems (DWSs) are all designed to harness or manage microbial communities for desired process outcomes (**Figure 1**). AS systems are designed to enrich for specific microbial lineages to remove nutrients, mostly organic carbon, phosphorus, and nitrogen species to prevent downstream receiving water eutrophication (Ardern and Lockett, 1914). AD processes are applied to various organic waste streams to recover biogas and other valuable products as a source of renewable energy (Angelidaki et al., 2011; Weiland, 2010). While biofilms in drinking water treatment plants (DWTPs) can leverage microbial communities for pollutant removal (Kirisits et al., 2019), DWSs apply physicochemical approaches to minimize microbial concentrations in tap water (Berry et al., 2006). All of these systems are the foundation of modern urban water infrastructure, and are intertwined through their dependence on adequately managing different microbial communities through a water-microbiome continuum (Raskin and Nielsen, 2019). As many of the key microorganisms underpinning these water cycles remain uncultured to date (Lloyd et al., 2018; Steen et al., 2019), our understanding of the microbial lineages involved in these engineered ecosystems have largely relied on cultivation-independent approaches.

      Tools for characterizing the structure and function of microbial communities have greatly expanded since the development of methods for surveying uncultivated microorganisms based on the use of the 16S ribosomal RNA (rRNA) gene (Fox et al., 1980; Lane et al., 1986; Woese and Fox, 1977). Building on the identification of specific 16S rRNA gene sequences, uncultivated microorganisms could be directly visualized using fluorescently

labelled oligodeoxynucleotide probes (DeLong et al., 1989) or microautoradiography (Giovannoni et al., 1988). Although advances in high-throughput sequencing technologies have allowed for large-scale 16S rRNA gene amplicon surveys instead of constructing labor-intensive clone libraries (Sogin et al., 2006; Caporaso et al., 2012; Kozich et al., 2013; Thompson et al., 2017), operationally defined taxonomical units cannot be linked to functional capabilities in mixed communities with fidelity, especially for poorly characterized lineages. Metagenomics is a path to solving this problem. The term metagenome is defined as the entire genetic content of the collective microbial genomes contained within a given environmental sample (Handelsman et al., 2002, 1998). Early metagenomic approaches cloned fragments of extracted microbial DNA into a host backbone (e.g., *E. coli*) to screen for particular sequences (Beja et al., 2000; Henne et al., 1999), functionally characterize novel enzymes through heterologous expression systems (Brady and Clardy, 2000; Rondon et al., 2000), or perform shotgun sequencing (Tyson et al., 2004; Venter et al., 2004). All of these approaches have the potential to link functional observations to phylogenetic identity, and therefore provide a powerful approach to characterize the ecology and evolution of uncultivated microorganisms.

Metagenomic sequencing approaches can be defined under two categories, read-based and assembly-based. Read-based metagenomic approaches map the resulting reads back to reference genomes to assess the taxonomic composition and functional potential of a mixed microbial sample (Quince et al., 2017b; Sharpton, 2014). However, read-based approaches heavily rely on reference databases, and therefore under-explored environments such as engineered water systems may not be well-represented in these databases. Additionally, this approach cannot directly link specific functions to individual microbial guilds or populations. Assembly-based approaches such as genome-resolved metagenomics extract individual genomes *de novo* from a mixed community through a process termed binning of assembled contigs or scaffolds (Albertsen et al., 2013; Tyson et al., 2004). The resulting genomes can be termed draft bins, metagenome-assembled genomes (MAGs), and/or population genomes, as they most often represent a mixed consensus sequence of closely related microbial cells (Bowers et al., 2017). These genomes can then be used as a foundation for integrated multi-'omics analyses using insights from metatranscriptomics, metaproteomics, and/or metabolomics that attempt to directly tie metabolic guilds and functional activities to specific microorganisms (Anantharaman et al., 2016; Woodcroft et al., 2018; Wrighton et al., 2012).

In recent years, there have been calls for microbiome science as a whole to shift away from mere descriptive studies and towards more mechanistic and predictive frameworks for designing and harnessing diverse microbial communities for desired beneficial outcomes (Alivisatos et al., 2015; Lawson et al., 2019). Successful adoption of predictive frameworks in engineered water systems could enable more economically and environmentally sustainable bioprocesses as demand for water and energy resources increases (Raskin and Nielsen, 2019; Sheik et al., 2014). To achieve this, an interdisciplinary approach combining theoretical microbial ecology and engineering principles must be applied. Genome-resolved metagenomic approaches have been widely adopted in water engineering applications to explore the diversity, metabolic potential, and spatiotemporal dynamics of microorganisms inhabiting water systems. Integrating multi-'omics approaches within an eco-systems biology framework is a promising approach for creating more predictable, controllable biotechnology applications for water microbiomes (Narayanasamy et al., 2015).

Here, we review existing work and prospects for multi-'omics applications in engineered water systems. We summarize pioneering work in genome-resolved metagenomic approaches over the last two decades in AS, AD, and DWS. We then describe how mechanistic modeling and time-series forecasting of microorganisms in water systems can integrate genome-resolved metagenomics to enhance their predictive capabilities. Finally, we highlight emerging 'omics technologies that will pave the way for remarkable advances not only in water systems but microbiome science as a whole.

# 1. GENOME-RESOLVED METAGENOMIC APPROACHES FOR UNCOVERING PRINCIPLES OF MICROBIAL ECOLOGY IN WATER SYSTEMS

Numerous molecular methods have been applied to engineered water systems to explore the diversity and ecophysiology of microorganisms, including 16S rRNA gene amplicon sequencing and in-situ hybridization approaches. However, these approaches alone are fundamentally limited in their ability to connect structure-to-function for any natural microbial community. Here, we review how early applications of molecular approaches have provided valuable insights into the microbial ecology of water systems. We then broadly summarize genome-resolved metagenomic approaches that span the methodologies of metagenomic binning, metatranscriptomics, and metaproteomics, and highlight their use in water engineering applications.

## *1A. Early Insights from Molecular Approaches*

The demonstratable success of molecular biology in the 1970s-80s in developing new methods for isolating, sequencing, and manipulating nucleic acid sequences naturally led microbial ecologists to apply these techniques to identifying uncultivated microbes in environmental samples (see extensive contemporary reviews in Amann et al., 1995; Olsen et al., 1986). Once the small subunit rRNA was identified as a suitable marker gene for phylogenetic analysis of taxa (Woese and Fox, 1977), several landmark studies demonstrated the feasibility of surveying the composition of natural microbial communities through comparing sequences of clone libraries and amplicons (Giovannoni et al., 1990; Schmidt et al., 1991). The utility of 16S rRNA gene profiling increased dramatically with the development of second-generation sequencing technologies initially with Roche 454 and subsequently Illumina, along with the development of related bioinformatics workflows (Bolyen et al., 2019; Callahan et al., 2016; Schloss et al., 2009). 16S rRNA gene amplicon sequencing surveys have been applied to infer the diversity and dynamics of specific microbial populations in full-scale WWTP configurations over space and time, revealing contributions of core microbiome members and rare taxa in these systems (Lawson et al., 2015; Saunders et al., 2016; Wu et al., 2019; Zhang et al., 2012). Based on 16S rRNA gene sequencing surveys, the Microbial Database for Activated Sludge (MiDAS) field guide was developed to connect functional information to manually curated taxonomical assignments for abundant microbial lineages in activated sludge (McIlroy et al., 2015), and was further updated to include anaerobic digestion systems (McIlroy et al., 2017).

The advent of 16S rRNA gene sequencing led to using radioactively-labeled oligonucleotide probes binding to rRNA structures to quantify the abundance of specific microbial populations (Giovannoni et al., 1988; Stahl et al., 1988). Panels of oligonucleotide probes were constructed for major subclasses of *Proteobacteria* and specific members of the Cytophaga-Flavobacter-Bacteroidetes (CFB) supergroup (Manz et al., 1996, 1992). Development of group-specific hybridization probes allowed for quantifying the relative abundance of methanogens in anaerobic digestion systems (Raskin et al., 1994a, 1994b), which were subsequently used to connect population dynamics to process performance parameters (Griffin et al., 1998; McMahon et al., 2001). Modifications to these methods including combining fluorescent in situ hybridization (FISH) with rRNA-targeted oligonucleotide probes and adding radioactively-labeled substrates through microautoradiography to visualize spatial co-localization and metabolic activities of microbial communities (Lee et al., 1999). These methods were particularly appealing for studying the dense, biofilm-like floc aggregates characteristic of activated sludge systems, such as quantifying and localizing filamentous and nitrifying bacteria within individual flocs (Mobarry et al., 1996; Wagner et al., 1994a). Subsequently, FISH-based methods combined with isotopes were used for understanding *in situ* activities (Amann, 1995; Wagner and Haider, 2012), and could also be used for verifying metabolic potentials of microorganisms made through hypotheses with reconstructed genomes (McIlroy et al., 2016).

Although both 16S rRNA gene sequencing and oligonucleotide probe applications have provided valuable insights into engineered water systems and continue to be used for novel insights, they carry significant limitations. 16S rRNA gene amplicon sequencing can be used as a high-throughput and low-cost tool to characterize microbial community dynamics through time and space; however, associated functional predictions cannot be captured *a priori* using this approach. Additionally, specific microbial traits underlying key biogeochemical transformations have been shown to be horizontally transferred and therefore not easily inferred from 16S rRNA gene identity (Anantharaman et al., 2018; Lawson and Lücker, 2018; McDaniel et al., 2020; Palomo et al., 2018; Parsons et al., 2020). The analysis of 16S rRNA gene amplicons remains complex due to conceptual difficulties surrounding the definition of OTUs (Edgar, 2018; Stackebrandt and Goebel, 1994), the propensity of OTU-clustering to generate misleading artifacts, limitations of cross-referencing short, noisy amplicon sequences across multiple studies, and the challenges of distinguishing closely related strains harboring similar 16S rRNA gene sequences. Some of these limitations have been addressed by constructing ecosystem-specific databases that can resolve amplicons at the species level due to using full-length 16S rRNA genes from long-read sequencing technologies (Nierychlo et al., 2020).

Read-based metagenomic approaches were initially used in engineered water systems to compare *in situ* species to existing reference genomes and measure the extent of similarity and population heterogeneity (Albertsen et al., 2011; MacIlroy et al., 2013). However, read-based metagenomic approaches depend on comprehensive reference databases to ensure accuracy of results. Although improvements in metagenomic sequencing efforts have increased the amount of reference genomes from environmental ecosystems (Nayfach et al., 2020; Parks et al., 2017), the fraction of high-quality genomes from engineered water systems still remains small (Hull et al., 2019; Nayfach et al., 2020). Therefore, substantial efforts are still needed to construct high-quality reference databases of genomes from anaerobic digestion, activated sludge, and drinking water system microbiomes. Although thorough methodological reviews and best practices in metagenomics exist elsewhere (Dick, 2018; Knight et al., 2018; Quince et al., 2017b), we provide brief overviews of existing and newly developed workflows and toolkits in genome-resolved 'omics to provide water engineers and scientists with the most recent advances and guidelines in a field subject to very rapid development. Lastly, we outline current technologies and approaches for applying integrated multi-'omics approaches to explore the dynamics and activities of microbial communities underpinning engineered water systems.

*1B. Genome-Resolved Metagenomics: From Individual Genomes to Large-Scale Reference Databases*

Genome-resolved metagenomic approaches apply *de novo* assembly to extract representative population genomes from a given system through metagenomic binning, and subsequently use draft bins with multi-'omics approaches to gain functional insights (**Figure 2**). The first objective of any genome-resolve approach is to assemble the representative members of a given community into MAGs. This process is termed "binning", in which short reads are *de novo* assembled into longer contigs, with groups of contigs hypothesized to arise from the same genomic context subsequently clustered based on sequence composition, differential coverage across time and/or space, or a combination of both parameters (**Figure 3**) (Albertsen et al., 2013; Alneberg et al., 2014; Dick et al., 2009; Sharon et al., 2013). These groups of "binned" contigs can be considered as working models of whole genomes, and are also referred to as population genomes, as the phylogenetic composition of assembled bins can be operationally defined by the minimum sequence identity of reads mapping back to assembled contigs (Jain et al., 2018; Olm et al., 2020). For example, bins are clustered and/or dereplicated at the 95% sequence identity threshold, and therefore a bin representing an "individual species" may actually be composed of multiple, closely-related strains that could not be assembled individually due to computational constraints (Jain et al., 2018; Olm et al., 2017). Draft bins can then be curated to verify uniform read coverage of all contigs (Bornemann et al., 2020; Eren et al., 2015), estimate approximate genome completion and contamination estimates by detecting universal single-copy genes (Parks et al., 2015; Seppey et al., 2019), dereplicate redundant sets of bins resulting from the use of binning results from multiple assemblies or binning algorithms (Olm et al., 2017; Sieber et al., 2018), or manually scaffold bins to close gaps and potentially recover completely circular MAGs (Chen et al., 2020; Lui et al., 2020).

After curation, taxonomical assignments and functional annotations are performed on all MAGs to connect structure to potential function. Currently, two main approaches exist for assigning taxonomy to draft genomes; an automated approach based on whole-genome sequence similarity against publicly available genomes (Chaumeil et al., 2019; Parks et al., 2018), and approaches based on clustering of ribosomal protein sequences followed by phylogenetic reconstruction with select reference genomes (Hug et al., 2016; Lee, 2019; McDaniel et al., 2019), both achieved through concatenating alignments of single copy marker genes. After assigning taxonomical classifications to genomes, specific metabolic pathways can be investigated based on gene presence and/or absence. Numerous pipelines exist for assigning functional annotations to draft genomes, and mostly differ in the underlying database(s) from which they draw functional annotations (Aramaki et al., 2019; Hanson et al., 2014; Seemann, 2014). Recently, several pipelines have been developed for exploring specific metabolic guilds within MAG datasets based on curated databases or marker profiles (McDaniel et al., 2019; Neely et al., 2020; Shaffer et al., 2020; Zhou et al., 2019). Furthermore, ecological and evolutionary insights can be made through comparative genomics approaches, such as hypotheses of niche differentiation based on gene content (Camejo et al., 2017; Flowers et al., 2009; Koch et al., 2015; Oyserman et al., 2016a; Skennerton et al., 2015; Speth et al., 2012), or analyzing microdiversity and population heterogeneity (Albertsen et al., 2011; Leventhal et al., 2018).

Some of the first demonstrations of genome-resolved metagenomics are from laboratory-scale wastewater enrichment bioreactors. The first representative genome of the polyphosphate accumulating organism (PAO) 'Candidatus Accumulibacter phosphatis' UW-1 was assembled from American and Australian enrichment bioreactors (Martín et al., 2006). Remarkably, both bioreactors happened to be enriched with similar Accumulibacter populations at the time of sequencing. Genomes of multiple Ca. Accumulibacter strains have since been assembled to date (Albertsen et al., 2016; Camejo et al., 2019; Flowers et al., 2013; Gao et al., 2019; Mao et al., 2014; Qiu et al., 2020; Skennerton et al., 2015), including high-quality genomes assembled from full-scale WWTPs (Law et al., 2016; Srinivasan et al., 2019), even though it remains to be isolated in pure-culture. The uncultivated bacterium *Kuenenia stuttgartiensis* (Strous et al., 2006) underlying anaerobic ammonium oxidation (anammox) was assembled from a complex bioreactor community, and subsequently other anammox species within the *Planctomycetes* have been assembled (Lawson et al., 2017; Speth et al., 2016). The genome of the nitrite-oxidizing bacterium *'Candidatus* Nitrospira defluvii' was assembled from an activated sludge enrichment bioreactor (Lücker et al., 2010), followed by other *Nitrospira* and *Nitrospina* species (Koch et al., 2015; Lücker et al., 2013). Remarkably, complete ammonium oxidation (comammox) was discovered by two groups through enrichment of *Nitrospira inopinata* from the pipe of a deep oil exploration well and two distinct *Nitrospira* species from a recirculation aquaculture system biofilter, respectively (Daims et al., 2015; Van Kessel et al., 2015).

Before the development of automatic binning methods, reconstruction of genomes from uncultured microbial lineages within a mixed population was mostly limited to a few abundant members to study their metabolism in detail, as the assembly and curation of any single MAG relied on largely manual approaches (Tyson et al., 2004). Albertsen et al. first demonstrated that MAGs of low abundant microorganisms within an activated sludge enrichment bioreactor could be captured using differential-coverage and composition based binning (Albertsen et al., 2013). Since then, numerous automatic binning algorithms that implement differential coverage profiles in addition to sequence composition signatures have been developed (Alneberg et al., 2014; Kang et al., 2015; Wu et al., 2014). Genome-resolved metagenomic surveys now routinely consist of aiming to assemble all representative species within a community for which medium- to high-quality MAGs can be reconstructed (Anantharaman et al., 2016; Bowers et al., 2017; Woodcroft et al., 2018). Recent examples of whole-community genome-resolved metagenomic analyses in engineered water systems include metagenomic assembly of *Ca.* Accumulibacter and 39 non-PAO "flanking" community members in a denitrifying enrichment bioreactor (Gao et al., 2019), investigating the impact of disinfection on microbial community structure and metabolism in a full-scale DWDS (Dai et al., 2019; Sevillano et al., 2020), nitrogen metabolism in drinking water reservoirs (Potgieter et al., 2020), and tracking the microbial community through a pilot plant for potable water reuse of wastewater (Kantor et al., 2019). Representative genomes from full-scale systems have only recently been constructed through large-scale sequencing efforts and curation of system-specific MAG databases, such as the collection of 1,600 MAGs from anaerobic digestion biogas reactors (Campanaro et al., 2020) and 2,045 MAGs assembled from a collection of 114 activated sludge full-scale WWTPs (Ye et al., 2020).

The use of MAG-based methods has allowed for the recovery of a large number of genomes from the uncultivated majority, however, current methods have several significant limitations. The most significant relates to the fractionated nature of MAGs assembled from short read data, which is a consequence of the limited read length and the inability to reconstruct regions containing complex repeat structures or share similarity between sub-species or strains, such as in ribosomal RNA operons (Chen et al., 2020; Meziti et al., 2021). This results in genomes that are comprised between ten to hundreds of contigs, which has implications for defining genome completeness and contamination with a high degree of confidence (Orakov et al., 2020). Furthermore, recovering genomes of individual strains from complex communities still remains a significant limitation of current methods, despite the highly sophisticated bioinformatics algorithms that have been developed for this problem (Olm et al., 2021; Quince et al., 2017a; Segata et al., 2012). All of these limitations are particularly acute in highly complex microbial communities such as those inhabiting full-scale activated sludge and anaerobic digester communities. According to the Minimum Information about a Metagenome-Assembled Genome (MIMAG), a high-quality MAG draft contains all three ribosomal rRNA genes, at least 18 tRNAs, and completion of >90% and contamination <5% calculated by the presence of single-copy marker genes (Bowers et al., 2017). A recent large-scale assembly and binning effort of publicly available metagenomes in the Integrated Microbial Genomes (IMG) system recovered less than 20% high-quality MAGs in the ~52,000 MAG (Nayfach et al., 2020). Approximately 3,400 of these MAGs were recovered from engineered water systems (**Figure 4**), and demonstrate that the majority of these MAGs are of medium-quality according to MIMAG standards. Although a large portion of these MAGs contain >90% completeness and <5% redundancy, they do not meet the high-quality standards due to missing rRNA and/or tRNA genes. As discussed below, obtaining high-quality MAGs is essential for ensuring accuracy of functional hypotheses and further validating these hypotheses with experiments. Using composite MAGs of medium-quality or low-quality can lead to inaccurate ecological inferences due to either inflating relative abundance calculations or missing genes conferring key biogeochemical transformations (Shaiber and Eren, 2019). In summary, MAG recovery from short read metagenomic data, while currently is routinely and relatively easily applied, is still subject to many caveats and substantive limitations.

### *1C. Integrated Multi-'Omics Approaches in Water Systems*

Integrated multi-'omics analyses consist of applying genome-resolved metagenomics in combination with one or more additional 'omics-based approaches, such as metatranscriptomics, metaproteomics, and/or metabolomics (**Figure 2**). Although each integrated approach can enhance our understanding beyond obtaining genomic sequences alone, these methodologies are currently less developed than metagenomic binning workflows, and may require more specific knowledge bases as well as system-specific methodological development to be performed effectively. It is important to note that in the case of metatranscriptomics and metaproteomics, the acquired data can be referenced directly back to recovered genome sequences. However, this is not the case for metabolomic data, which requires known or putative metabolic reaction pathways to be known, or predicted, in order to link detected compounds with the corresponding gene products encoding the cognate enzymes.

Therefore, in both genome-resolved metatranscriptomics and metaproteomics approaches, short cDNA reads or peptide sequences are competitively mapped against a collection of genomes to identify which genes are actually being expressed and to provide estimates of relative expression levels (Shakya et al., 2019; Wilmes and Bond, 2009; Woodcroft et al., 2018). Successfully conducting metatranscriptomics and metaproteomics is substantially more complex than encountered in metagenomics. The extraction and handling of RNA and protein are complicated by the fact that both molecules are far less stable than DNA (Moran et al., 2013). Furthermore, because approximately 80-90% of the RNA in a given cell is mostly ribosomal RNA (Westermann et al., 2012), then the remaining ~10-20% fraction of the reads from total RNA sequencing are from protein-coding genes. At the whole community level, this can result in noise-effected or false-negative calls for genes in all but the most abundant genomes. Accordingly, rRNA-depletion procedures are best applied following the extraction of total RNA (Culviner et al., 2020; He et al., 2010b; Stewart et al., 2010). Collectively, these issues require complex bench workflows, requiring a high degree of molecular biological knowledge and skill. At the data analysis level, because sequencing-based gene expression estimates are generally relative, it can be challenging to understand the relationship between changes in gene expression and fluctuations in taxon abundance. Therefore, co-extraction of DNA, RNA, and/or protein from the same aliquot is recommended, combined with recovery of MAGs from the DNA fraction in order to accurately capture the cognate genomic and gene sequences underpinning RNA-seq data (Roume et al., 2013a, 2013b).

Traditionally, metatranscriptomics analyses employ either gene-centric or genome-centric approaches, both of which depend on identifying highly, differentially, or co-expressed genes (Shakya et al., 2019). Gene expression dynamics of Accumulibacter were first elucidated using oligonucleotide microarrays (He et al., 2010a), with a more detailed metabolic reconstruction of this lineage generated using time-series RNA sequencing (Oyserman et al., 2016b), and ecological roles for co-existing Accumulibacter clades inferred through comparative genome-resolved metatranscriptomics (McDaniel and Moya et al., 2020; Wang et al., 2020). Genome-resolved metatranscriptomics has also been used to demonstrate activities of specific genes, such as a complete denitrification pathway in Accumulibacter clade IC (Camejo et al., 2019), and extracellular electron transfer activities of a wastewater-fed microbial full cell community (Ishii et al., 2013) and anammox bacteria (Shaw et al., 2020) . Expanding beyond the detailed study of a single microbial lineage, genome-resolved metatranscriptomic approaches have been applied to whole communities for which there is sufficient read coverage (Woodcroft et al., 2018) to infer putative interactions and connect gene expression patterns to overall ecosystem functioning. Lawson et al. (2017) applied genome-resolved metatranscriptomics to 17 assembled population genomes from an anammox bioreactor to infer interactions between heterotrophic and annamox bacteria. Scarborough et al. (2018) integrated genome-resolved metatranscriptomics and thermodynamic analysis to elucidate microbial interactions contributing to medium-chain fatty acid production of an anaerobic microbiome. Hao et al. (2020) reconstructed 182 MAGs from two full-scale anaerobic digestors and applied genome-resolved metatranscriptomics to explore gene expression programs in response to short-chain fatty acid stimulation.

Proteomics methods couple liquid chromatography (LC) with tandem mas-spectrometry (MS) to profile the protein content of a single organism or microbial community (Lipton et al., 2002; Wilmes and Bond, 2006). Overall, genome-resolved metaproteomics approaches as a whole have not been applied as frequently as metatranscriptomics. This is most likely due to both the technical difficulties in high-throughput mass spectrometry methodologies and development of appropriate bioinformatics pipelines for competitively mapping short peptide sequences to a collection of assembled genomes. However, landmark studies have applied proteomics to whole communities to hone in on the functional activities of dominant lineages (Lo et al., 2007; Ram et al., 2005; Roume et al., 2015a; Woodcroft et al., 2018). In engineered water systems, shotgun proteomics have been applied to enrichment bioreactors to reconcile established metabolic models and identify strain variants of Accumulibacter (Wilmes et al., 2008) and analyze complexomes of anammox bacteria (de Almeida et al., 2016).

Fully integrated studies across the genomic, transcriptomic, proteomic, and even metabolomic landscapes have only recently been produced for engineered water systems. Methodologies for sequential extraction of DNA, RNA, protein, and metabolites from a single microbial community sample have enhanced the reproducibility of integrated 'omics measurements (Roume et al., 2013a, 2013b). An integrated comparative 'omic analysis using metagenomics, metatranscriptomics, and metaproteomics of the anoxic tank of a full-scale WWTP was carried out to construct community-wide metabolic networks (Roume et al., 2015b). Metabolomics paired with genomic, transcriptomic, and proteomic data was used to understand lipid accumulation in WWTPs and the role of *'Candidatus* Microthrix parvicella' (Muller et al., 2014). Genome-resolved metagenomic, metatranscriptomic, and metaproteomic approaches were applied to a partial-nitration anammox lab-scale bioreactor to explore operational parameters and microbial interactions affecting nitrogen removal (Wang et al., 2019). Recently, an integrated time-series genome-resolved metagenomics, metatranscriptomics, metaproteomics, and metabolomics investigation of full-scale WWTP microbial community recovered 1,364 MAGs and demonstrated patterns of resistance and resilience to natural environmental perturbations (Herold et al., 2020). As technologies and bioinformatics algorithms improve, integrated multi-'omics approaches have the potential to enable powerful ecological insights into engineered water microbiomes and create the foundation for more predictable and controllable systems (Muller et al., 2013).

## 2. PAIRING 'OMICS-FUELED INSIGHTS WITH MODELING APPROACHES

Obtaining genome sequences for uncultivated microbial lineages and integrating spatiotemporal metatranscriptomic and metaproteomic data has become increasingly commonplace in research conducted by microbial ecologists and water engineers. Looking ahead, effective prediction and control of microbial communities for desired outcomes in engineered water systems is likely to rely in part on the ability to integrate vast amounts of 'omics data into process modeling frameworks. In this section, we describe two overarching aims of broad modeling approaches used to *1*) interrogate metabolic networks and *2*) predict the temporal dynamics of complex communities. For metabolic modeling approaches, we summarize how emerging techniques in metabolomics and stable isotope probing are being used in engineered water systems. For modeling population dynamics, we briefly review how 16S rRNA gene sequencing approaches have been used for time-series forecasting in water engineering, and how these studies can be strengthened by applying genome-resolved multi-'omics approaches.

### *2A. Metabolic Modeling with Pathway Flux, Stable Isotope Probing and Metabolomics*

The metabolic capabilities of a microorganism depend on the available reaction networks, putative intracellular storage compounds, and environmental conditions. The past few decades have seen major developments in metabolic modeling methods, fueled by more quantitative data from the application of different 'omics approaches as well as increased computational power. These approaches have ranged from "simple" stoichiometric models to complex spatiotemporal, multi-scale simulations of ecosystems (Bauer et al., 2017). Annotation and generation of genome-scale stoichiometric models is becoming a more automated process with different tools available for binning, functional annotation, and metabolic model generation (Hamilton and Reed, 2014; Machado et al., 2018). Overall, these predictions are based on 1) the metabolic reaction networks of different microorganisms, which can be derived from annotated genomes, at least within the limits of known biochemistry (Feist et al., 2009; Hamilton and Reed, 2012) 2) expression levels of the required enzymes and their activities (Hamilton et al., 2017; Molenaar et al., 2009), and 3) the kinetics of the extracellular environment and metabolite concentrations (Herold et al., 2020).

The exploration of metabolic network functionality has been enabled by stoichiometric modeling approaches, building on the availability of whole-genome sequences, namely flux balance analysis (FBA) (Orth, Thiele and Palsson, 2010) or flux variability analysis (FVA) toolkits (Gudmundsson and Thiele, 2010). These approaches were first developed for steady-state systems and were able to predict biological phenomena such as overflow metabolism in a single organism (Basan et al., 2015), as well as cross-feeding consortia (Stolyar et al., 2017; Carlson et al., 2018). Extensions to these approaches have been made to more dynamic and cyclic conditions to correctly predict storage accumulation and depletion cycles in different organisms (Reimers et al., 2017). For example, applications of FBA and FVA in engineered water systems include the study of *Ca.* Accumulibacter to highlight the flexibility of anaerobic metabolism due to intracellular accumulation of different polymers (Guedes da Silva et al., 2020). Genome-scale models can also be used to understand the relationship between thermodynamics and metabolism, such as modeling an anaerobic syntrophic coculture (Hamilton et al., 2015). Weinrich et al. (2019) developed a framework to incorporate FBA into the IWA Anaerobic Digester Model No.1 to predict thousands of intracellular metabolite fluxes of individual species in anaerobic digesters, providing an opportunity to couple process modeling with multi-omics datasets. Integrated 'omics approaches can also aid in improving genome-scale metabolic models and thermodynamic analyses. Scarborough et al. used genomic and transcriptomic data to understand interactions and substrates of an anaerobic microbiome producing medium-chain fatty acids (Scarborough et al., 2018). They subsequently used time-series transcriptomic data and metabolomics to understand the environmental conditions favoring medium-chain fatty acid production (Scarborough et al., 2020b) and construct unicellular and guild-based metabolic models of the community (Scarborough et al., 2020a).

Stable isotope probing (SIP) and tracing approaches have been applied to highly enriched and complex microbial communities to identify metabolically active populations (**Figure 5**). Metabolically active populations can be further taxonomically and functionally characterized by pairing heavy labelled $^{13}$C isotopes with metagenomic sequencing (Dumont and Murrell, 2005; Radajewski et al., 2000) or metaproteomic sequencing (Jehmlich et al., 2008; Von Bergen et al., 2013). Incorporation of the $^{13}$C isotope into a given carbon source then allows for $^{12}$C and $^{13}$C labelled fractions to be resolved by density-gradient centrifugation (for DNA/RNA-SIP) and separately characterized by amplicon or shotgun metagenomic sequencing approaches (Chemerys et al., 2014; Chen and Murrell, 2010; Coyotzi et al., 2016; Neufeld et al., 2007; Saidi-Mehrabad et al., 2013), or by mass-spectrometry approaches for protein-SIP (Sachsenberg et al., 2015). Genome-resolved DNA-SIP approaches can be achieved by exploiting abundance patterns between $^{12}$C and $^{13}$C density fractions and applying differential coverage binning to metagenomic samples (**Figure 5**). Genome-centric SIP approaches were demonstrated to elucidate populations that could degrade oleate (Ziels et al., 2018) and butyrate (Ziels et al., 2019) in anaerobic digesters. Insights into novel syntrophic acetate oxidizers in full-scale anaerobic digesters were also obtained by coupling genome-resolved metagenomics with metaproteomics-SIP (Mosbæk et al., 2016).

Isotopic tracers can also be used in metabolic flux analyses to measure the *in vivo* flux of metabolites in individual organisms (Ando and García Martín, 2019; Matsuoka and Shimizu, 2014) or cocultures (Gebreselassie and Antoniewicz, 2015). Quantifying metabolic flux changes under different environmental conditions using isotopic tracers can provide insights into metabolic regulation (Antoniewicz, 2015; Long and Antoniewicz, 2019; Nielsen, 2003). Lawson et al. used a combination of time-series $^{13}$C and $^{2}$H isotopes, metabolomics, and isotopically nonstationary metabolite flux analysis to resolve the central carbon metabolism of a highly enriched anammox bioreactor community (Lawson et al., 2020). Although metabolic flux analyses are difficult to perform for complex microbial communities, advances in assigning individual peptides and metabolites to a given species has started to pave the way for metafluxomics (Beyß et al., 2019; Ghosh et al., 2014).

Recent advances in high-throughput mass spectrometry-based technologies have allowed for metabolomics approaches to shift from targeted characterization of known metabolites to untargeted surveys of numerous metabolites and other small molecules (Schrimpe-Rutledge et al., 2016). Although both targeted and untargeted strategies have their shortcomings and challenges, untargeted approaches enable discovery-based studies that can be subsequently tested with targeted approaches (Cajka and Fiehn, 2016). Metabolomics can be combined with other multi-omics and modeling approaches to compare results from the level of gene regulation to metabolite production of a given system or process (Herold et al., 2020; Lawson et al., 2020; Muller et al., 2014; Roume et al., 2015a; Scarborough et al., 2020a). Recent advances in methods for simultaneous, nondestructive extraction of DNA, RNA, and metabolites from a given sample have paved the way for these types of intrasample datasets (Roume et al., 2013a, 2013b). These multi-omics approaches integrated with metabolic modeling approaches can greatly enhance our ability to predict the environmental conditions that are favorable for desired outcomes in engineered water systems.

### 2B. Genome-Resolved 'Omics Applied to Time-Series

Time-series data have been applied to numerous ecosystems to unravel the spatiotemporal dynamics of microbial communities in the context of environmental parameters (Shade et al., 2013). For example, longitudinal studies have been used to identify marine seasonal microbial community dynamics (Gilbert et al., 2012), community succession patterns in the developing infant gut (Palmer et al., 2007), and characterize patterns of microbial community recovery and resilience after ecosystem disturbances (Shade et al., 2012). Microbial community time-series have also been used to address important questions in engineered water systems such as community assembly in full-scale wastewater treatment plants (Griffin and Wells, 2017; Lee et al., 2015), population dynamics and immigration of anaerobic digestion communities (Griffin et al., 1998; Kirkegaard et al., 2017), and monitoring bacterial community migration dynamics in drinking water distribution systems (Boers et al., 2018; Pinto et al., 2014). However, most time-series resolved data have been generated using 16S rRNA gene amplicon sequencing surveys, which have significant shortcomings. Assigning metabolic functions to 16S rRNA gene based OTUs is not entirely intuitive for most guilds, such as heterotrophic bacteria that span diverse phyla (Frigon and Wells, 2019; Marques et al., 2017). In addition to some taxa in engineered water systems remaining unclassified, 16S rRNA gene based OTUs usually cannot be resolved more specifically than the genus-level (Kirkegaard et al., 2017).

Although long-read sequencing technologies (discussed below) can improve the resolution of resulting 16S rRNA gene based OTUs (Karst et al., 2021; Nierychlo et al., 2019), time-series data integrating metagenomics and integrated genome-resolved 'omics methods can be a powerful approach (Faust et al., 2015). Metagenomes collected over time can be used to improve binning efforts based on differential coverage profiles (Kang et al., 2015), and subsequently identify metabolic guilds and abundance patterns (Linz et al., 2018). If full-length 16S rRNA gene sequences are retrieved from assembled MAGs, defined microbial guilds based on metabolic reconstructions can be connected to 16S rRNA amplicon sequencing surveys of a given system (Nierychlo et al., 2019; Petriglieri et al., 2020; Singleton et al., 2020). Pérez et al. assembled 173 MAGs from a full-scale WWTP and assessed associated functional potential and the dynamics of rRNA gene operons during operational disturbances (Pérez et al., 2019). In conjunction with 16S rRNA amplicon sequencing, assembled MAGs were used to characterize functional profiles between operational periods (Pérez et al., 2019). Integrated genome-resolved 'omics approaches can also be applied in longitudinal analyses to understand community dynamics and functional profiles. Time-resolved 'omics approaches were recently applied to a full-scale wastewater treatment plant and demonstrate the power of fully integrated 'omics methodologies (Herold et al., 2020; Martínez Arbas et al., 2021).

Other promising developments have combined time-series 'omics data with biogeochemical models to predict the fate of nutrients in environmental systems (Louca et al., 2016; Reed et al., 2014). Such ecosystem level models have been constructed based on the abundance of genes in metagenomes (Reed et al., 2014), as well as transcript and protein abundances (Louca et al., 2016). The underlying model structure of such gene-centric modeling approaches is amenable to their adaptation into existing water engineering models (e.g. IWA, ASM, and ADM). However, 'omics data is inherently compositional, and therefore novel methods are needed to accurately infer relative or absolute taxa abundances from 'omics data so the stoichiometry can be maintained within process models. While some recent methods show promise, such as using internal quantitation standards for metagenomes (Hardwick et al., 2018), this task is not trivial due to differences in cell size, composition, biomass concentration, and complexity of biomass morphology.

## 3. EMERGING TECHNOLOGIES IN 'OMICS AND WATER ENGINEERING

Over the last few years, advances in sequencing technologies and novel methodologies have emerged that have direct implications for improving genome-resolved metagenomic insights. Although there have been various novel, cultivation-independent inventions for broadly characterizing microbial communities, we focus on two fields that have exhibited tremendous growth and opportunity for leveraging multi-'omics approaches specifically in engineered water systems. First, we summarize how improvements in the quality and substantial decrease in costs of long-read sequencing technologies have allowed for obtaining complete, closed, and high-quality reference genomes of microorganisms from mixed communities. We then highlight emerging cell-sorting techniques that can be applied in conjunction with multi-'omics for novel functional insights.

### *3A. High-Quality Genomes Enabled by Advances in Sequencing Technologies*

Although next-generation sequencing technologies have greatly enhanced high-throughput surveys of microbial communities, the emergence of so-called "third-generation" sequencing technologies using single molecule sequencing to produce longer reads are poised to revolutionize *de novo* assembly of complex genomes and metagenomes (Lee et al., 2016). Primarily, Pacific Biosciences (PacBio) and Oxford Nanopore Technologies (ONT) have emerged as the main technologies for long-read metagenomic surveys of microbial communities (Arumugam et al., 2021, 2019; Bertrand et al., 2019; Moss et al., 2020; Singleton et al., 2021; Somerville et al., 2019; Stewart et al., 2019). The greatly enhanced read length (routinely greater than 10k bp) presents substantial advantages in the process of genome assembly, by permitting the successful reconstruction of genome regions containing complex repeats, such as multiple intragenomic rRNA operons. Such repeat regions typically contribute to the high degree of fragmentation observed in assemblies constructed with short read data. In the case of gDNA sequencing from culture isolates, the use of long read technologies will typically result in the generation of a complete closed genome, represented a single, continuous sequence construct (Frank et al., 2018).

The major disadvantage of current long read methods relates to reduced sequence quality compared to short read sequencing, sometimes cited as high as 10-30% (Amarasinghe et al., 2020), whereas short read sequencing is likely to have an error rate no more than 1%. Rapid progress in sequencing chemistry has led to reduction in these error rate, with some reports of error rates <5% error using ONT sequencing and <1% with PacBio (as discussed by Amarasinghe et al., 2020). The impact of high error rates can readily be seen in protein coding sequences, where the occurrence of frame-shift errors (insertion/deletion, or indels) can be up to 80 times higher than that observed in Illumina sequences (Huson et al., 2018). Mitigating the impact of error rates requires the application of complex data correction procedures (Morisse et al., 2020), either self-correction of the long read data themselves, approaches that make use of complementary higher-accuracy short reads, or correction via the use of sequence alignment to reference sequence databases (Huson et al., 2018). At the time of writing, the development of base-calling methods and new error-correction procedures is a highly active research topic (Morisse et al., 2019).

In the context of water engineering, Arumugam and colleagues (Arumugam et al., 2021, 2019) demonstrated the recovery of complete genomes (single chromosome), and in some cases, circularized, closed genomes, of the most abundant community members from an enrichment bioreactor (22 in total, including key PAO, GAO, and filamentous species, among others; of which 10/22 were circularized). Recently Singleton et al. analyzed approximately 1Tbp of Nanopore data generated from 23 full scale WWTPs in Denmark, and obtained 1,083 high quality MAGs from 581 distinct prokaryotic species, including 57 with circularized closed genomes (Singleton et al., 2021). Notwithstanding these advances, there remain many emerging problems that will need attention with long read MAG recovery, such as the need for long read specific binning procedures (Singleton et al., 2021) and the impact of sequencing error rate on coding sequence quality (Arumugam et al., 2020).

In addition to technology advancements that directly allow for longer-reads from single molecules, novel, synthetic long-read approaches such as chromosome conformation capture (also referred to proximity ligation), linked-reads, and optical mapping (Amarasinghe et al., 2020) also have enormous potential for applications in water systems. Chromosome conformation capture (3C) methods apply formaldehyde cross-links prior to DNA lysis and extraction to create interaction points between nearby loci (Dekker et al., 2002). Crosslinked chromatin is then digested with specific restriction enzymes to obtain long-range interaction pairs, and used to analyze the contact frequencies of these pairs (Lieberman-Aiden et al., 2009; Sati and Cavalli, 2017). Originally, variants of 3C technologies were developed to study the three-dimensional organization of chromosomes and genome structures (Hsieh et al., 2015; Rao et al., 2014). Recently, this method has been adapted for generating genomes from semi-complex and complex microbial communities, such as assembling a yeast hybrid from a beer sample, population genomes from the human gut, and used in addition to standard Illumina shotgun sequencing to obtain ~900 genomes from the cow rumen (Burton et al., 2014; Marbouty et al., 2014; Press et al., 2017; Smukowski Heil et al., 2018; Stewart et al., 2018). Since this method produces linkages between genetic content within the same cell, this approach can also produce associations between antibiotic resistance genes (ARGs), plasmids, and mobile elements to the host genome with fidelity. This was recently demonstrated in a microbial community from a full-scale WWTP to link ARGs and plasmids to specific populations (Stalder et al., 2019), while also highlighting the challenges associated with implementing Hi-C approaches in complex and poorly studied microbial communities.

Although this review mainly covers genome-resolved approaches, long-reads have also been applied to improve amplicon sequencing surveys. Briefly, these high-throughput pipelines generate thousands to millions of full-length 16S rRNA genes (or any other amplicon of choice) from mixed microbial communities without primer-bias (Karst et al., 2021, 2018). The recently updated release of the MiDAS Field Guide with full-length 16S rRNA references allows for species-level classifications of microorganisms in activated sludge and anaerobic digestion systems (Nierychlo et al., 2019). Singleton et al. also demonstrated the ability to connect structure to function with fidelity by comparing full-length 16S rRNA sequences from large-scale microbial surveys to corresponding high-quality MAGs (Singleton et al., 2020). Full-length 16S rRNA sequences from high-quality MAGs were also used in Singleton et al. 2020 to construct novel FISH probes to visualize the dynamics of specific and understudied microorganisms in full-scale WWTPs (Singleton et al., 2020). Advances in long-read sequencing technologies will allow for the rapid recovery of high-quality genomes from complex microbial communities to use in conjunction with detailed experiments (**Figure 6**).

### *3B. 'Omics Approaches Enabled through Cell-Sorting Techniques*
Cell-sorting methodologies involve the separation of cells from a complex sample into simpler subsets for various downstream purposes. Sorting can be accomplished through various strategies, such as fluorescently or physically tagging cells with specific probes or dyes, randomly dividing groups of cells, or scanning cells for the presence of specific polymers or storage products (Malmstrom and Eloe-Fadrosh, 2019). Advances in cell-sorting technologies have allowed for both fundamental discoveries and applied practices in engineered water systems. Flow cytometry has been commonly used to monitor microbial communities in DWDS (Besmer et al., 2014; Douterelo et al., 2014), sometimes in conjunction with next-generation sequencing approaches (Prest et al., 2014). In relation to 'omics-based approaches, cell-sorting methods allow for downstream applications such as sequencing of simpler "mini-metagenomes" of a complex community, performing targeted cultivation of specific microbial lineages, and characterizing individual microbial cells exhibiting specific traits or physiologies (**Figure 6**).

Complex microbial communities can be subjected to cell-sorting techniques prior to DNA extraction and sequencing, leading to sequencing a "mini-metagenome" (Ji et al., 2017). This could entail randomly dividing cells by microfluidic sorting (Yu et al., 2017) or nonrandomly by fluorescently-activated cell sorting (FACS) (Schulz et al., 2018). Single amplified genomes (SAGs) are obtained through non-specific staining prior to sorting and screened for sequencing (Clingenpeel et al., 2014; Rinke et al., 2013). However, genomes from single-cell sequencing efforts are usually incomplete, contain high amounts of contamination, and can be expensive to obtain for low abundant microorganisms (Rinke et al., 2013; Swan et al., 2011; Xu and Zhao, 2018). Targeted approaches employ FISH probes to stain specific microbial lineages prior to FACS (Kalyuzhnaya et al., 2006), which can also allow access to low abundant microorganisms (Podar et al., 2007; Tan et al., 2019; Yilmaz et al., 2010). Improvements in FISH-based methods for staining specific microbial clades have led to higher quality genomes reconstructed from these efforts (Grieb et al., 2020). Intriguingly, a reverse genomics-based approach implementing engineered antibodies to capture specific microorganisms from a complex community was applied to culture members of the Saccharibacteria (TM7) (Cross et al., 2019). Live cell sorting was applied to marine nitrite-oxidizing enrichments to isolate novel *Nitrospinae* strains and investigate their nitrite affinity and metabolic differences (Mueller et al., 2020). Targeted cell sorting through fluorescent labeling combined with single-cell metagenomics was used to enrich for low abundant Chloroflexi species from a full-scale wastewater treatment plant (Dam et al., 2020).

In addition to targeting specific microorganisms based on phylogenetic identity, microbial populations can be screened and selected based on specific metabolic activities. Raman microspectroscopy methods quantify the scattering of light given off by chemical bonds to generate a spectral profile of specific biomarkers at the single-cell level (Huang et al., 2004; Neufeld and Murrell, 2007). Raman spectroscopy can be combined with stable isotope probing (SIP) and FISH methods to connect the phylogenetic identity of cells exhibiting specific metabolic activities (Huang et al., 2007). These methods have been used to retrieve cells based on specific functions before sequencing (Lee et al., 2020) or confirming metabolic activities of microorganisms based on genome sequences (Fernando et al., 2019). Raman cell-sorting in conjunction with mini-metagenomics was applied to a mouse gut microbial community to identify commensals that incorporate mucus-derived monosaccharides and rationally design a probiotic microbial consortia (Pereira et al., 2020). In engineered water systems, Raman microspectroscopy-based methods have been particularly used to identify polyphosphate accumulating organisms (PAOs) based on quantifying intracellular storages of polyphosphate (Fernando et al., 2019). In addition to the model PAO *Ca.* Accumulibacter, members of the actinobacterial genus *Tetrasphaera*, multiple species of *Dechloromonas,* the filamentous microorganism *Ca.* Microthrix, and the novel *Ca.* Methylophosphatis have been shown to incorporate polyphosphate intracellularly using Raman microspectroscopy imaging and quantification methods (Fernando et al., 2019; Petriglieri et al., 2021, 2020; Singleton et al., 2020). Enzymatic tagging using click-chemistry based approaches followed by cell sorting can be used to enrich low abundant populations for targeted metagenomic analyses (Sakoula et al., 2021). Bioorthogonal non-canonical amino acid tagging (BONCAT) is one such substrate analogue probing technique that labels newly translated proteins via azide-alkyne click-chemistry, and can be used to selectively sort active microorganisms from complex communities and environments (Couradeau et al., 2019; Hatzenpichler et al., 2016; Reichart et al., 2020). Advances in cell-sorting methods provide powerful approaches to apply identity- or function-based tagging to specific microbial populations in a complex community such as engineered water systems. These approaches can be used to construct mini-communities or synthetic consortia designed to perform a desired function, allowing for more controllable and predictable dynamics and outputs (Lawson et al., 2019; Pereira et al., 2020).

**4. CONCLUSIONS**
In the release of *Microbial Ecology of Activated Sludge* in 2010 (Seviour and Nielsen, 2010), with extensive field updates to the 1999 publication of *The Microbiology of Activated Sludge* (Seviour and Blackall, 1999), the editors noted that the greatest future challenge would be "to communicate with engineers to convince them that all this elegant science can be applied productively to improve how they design and operate their activated sludge systems" (Seviour and Nielsen, 2010). In the decade since, numerous breakthroughs and technological advancements in water-related microbiome science as a whole have stemmed from the desire to unravel the ecology and evolution of the microorganisms underpinning these systems. Clearly, the water engineering community is more convinced than ever that state-of-the-art molecular and 'omics tools can provide valuable insights into the microbiology of their systems. However, applying these insights to improve the design and operation of engineered water systems remains a grand challenge. Ultimately, successfully harnessing engineered water microbiomes for improved process design cannot be accomplished without effectively bridging engineering principles with microbial ecology theory (Lawson et al., 2019). Advances in genome-resolved multi-omics approaches provide an exciting new avenue for water engineers and microbial ecologists to tackle these interdisciplinary grand challenges.

**ACKNOWLEDGEMENTS**
This review resulted from materials and discussions held during the *Prospects for Multi-Omics Methods in Water Engineering* pre-conference workshop of the 8th International Water Association Microbial Ecology and Water Engineering Conference (MEWE2019, 17-20 November 2019, Hiroshima Japan). We thank Akiyoshi Ohashi, Futoshi Kurisu, and Kengo Kubota for their support as part of the organization of MEWE2019, and the Japanese Society of Water Environment (JSWE) for financial support of the event. We thank Emilie Muller (University of Strasbourg/CNRS) and Yu Ke (Peking University, Beijing) for preliminary discussions on the workshop content.

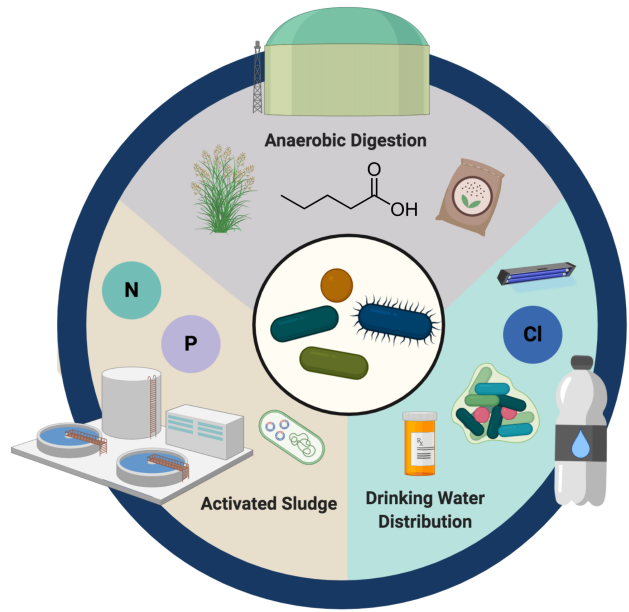

**Figure 1. Engineered water systems unified through microbial ecology:** engineered water systems are unified through microbial ecology in that they all aim to manage diverse microbial communities for desired outcomes. *Anerobic Digestion* (AD) systems apply microbial consortia to produce valuable bioenergy products from waste streams. *Activated Sludge* (AS) wastewater treatment systems harness the unique metabolic physiologies of specific microorganisms to remove nutrients such as nitrogen and phosphorus. *Drinking Water Distribution Systems* (DWDS) aim to manage microbial pathogens and biofilms to provide safe, clean drinking water. Created with BioRender.com.

**Figure 2**

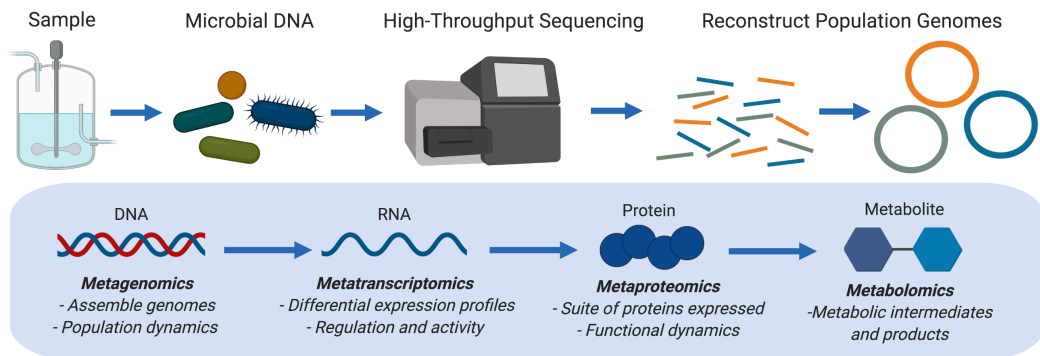

**Figure 2. Genome-resolved multi-'omics approaches**: genome-resolved approaches can be applied individually at the genomic, transcriptomic, and proteomic levels, or in conjunction with another. Microbial DNA is extracted from the environment of interest and used for high-throughput shotgun sequencing. Short reads are assembled into longer contigs, which are used to bin into population genomes based on sequence composition and differential coverage through space and/or time. Typical workflows for binning and downstream analyses of assembled genomes are outlined in Figure 3. Population genomes from short-read data usually represent a consensus sequence of closely related strains in a sample. Created with BioRender.com.

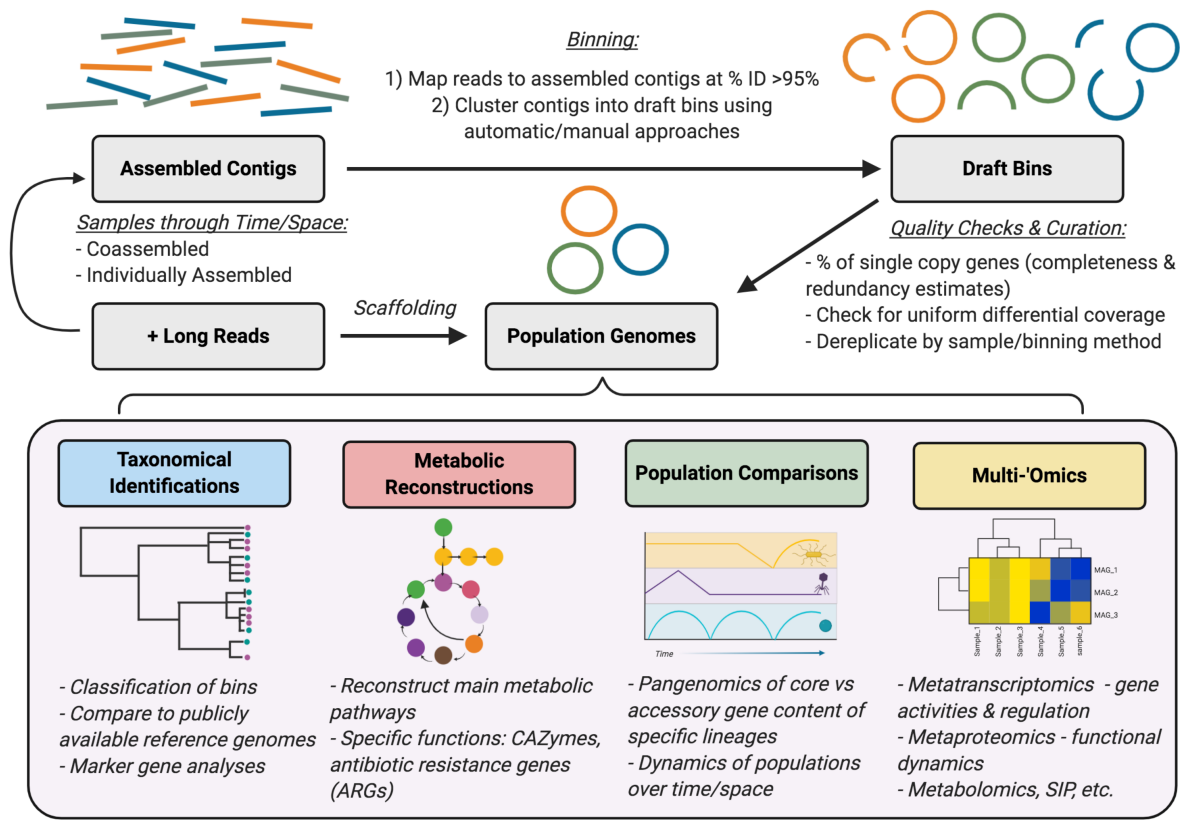

**Figure 3. Workflow for binning population genomes**: from sequenced metagenomes, samples are either assembled individually or coassembled together into longer contigs. These contigs are binned into draft bins by mapping reads back to assembled contigs at a percent identity >95%, and clustered by differential coverage profiles over space and/or time and/or using nucleotide signatures. These draft bins are then quality checked and curated based on the presence of single copy genes, checked for uniform differential coverage across all contigs, and dereplicating by sample or binning method. The resulting set of population genomes represent non-redundant, "species"-level bins that can be recovered from the community. Long reads can be incorporated either during initial assembly process with polishing tools to improve contigs lengths, used for scaffolding individual bins, or potentially as the basis for genome recovery (not shown). Population genomes can then be used for a variety of analyses to understand the given ecosystem – phylogenomic comparisons, metabolic reconstructions, dynamics of specific populations, and/or integrating with other multi-'omics approaches. Created with BioRender.com.

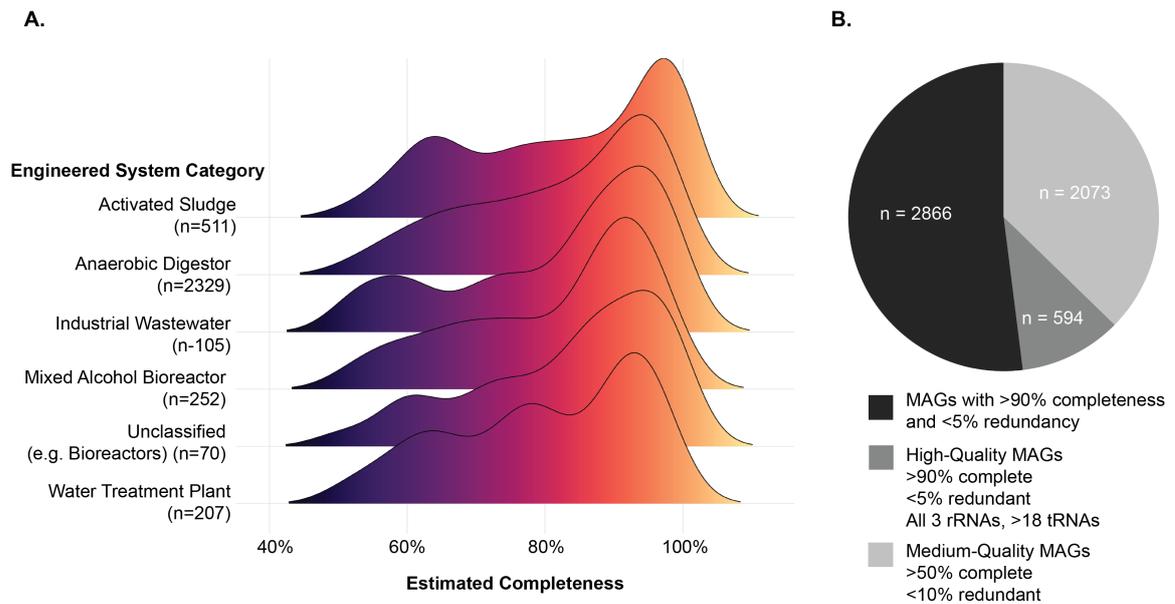

**Figure 4. Quality of MAGs from engineered water systems:** genome information from the Genomes from Earth's Microbiomes (GEM) catalog (Nayfach et al., 2020). From this dataset, genomes were subset by ecosystem and ecosystem category to only include genomes from engineered water systems (activated sludge, anaerobic digestion, drinking water systems, and lab-scale bioreactors simulating these processes). From 52,515 assembled MAGs, 3,474 MAGs were determined to be recovered from engineered water systems. **A.** Distribution of estimated completeness of all MAGs from different environmental categories of engineered water systems. **B.** Distribution of MAGs from engineered water systems that full under the categories high-quality, medium-quality, and genomes that are >90% complete and <5% redundant but do not contain all 3 rRNA genes and at least 18 tRNAs, as defined by the MIMAG standards (Bowers et al., 2017).

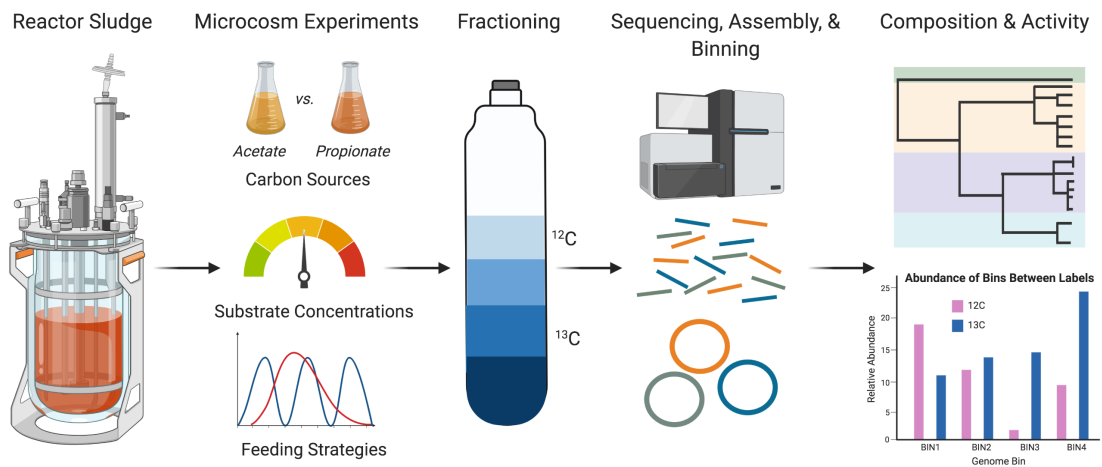

**Figure 5. DNA Stable Isotope Probing (SIP) paired with genome-resolved metagenomics**: DNA stable isotope probing (SIP) can identify active populations in a mixed community. Reactor sludge is inoculated into microcosms used for incubation experiments such as testing differences between carbon sources, substrate concentrations, or feeding strategies on community dynamics. Labelled DNA is density centrifuged and separated between light and heavy labelled DNA. The fractions are sequenced and assembled. Individual assembly of metagenomic contigs and differential coverage binning can then yield insights into the activities of specific populations between conditions. Created with BioRender.com.

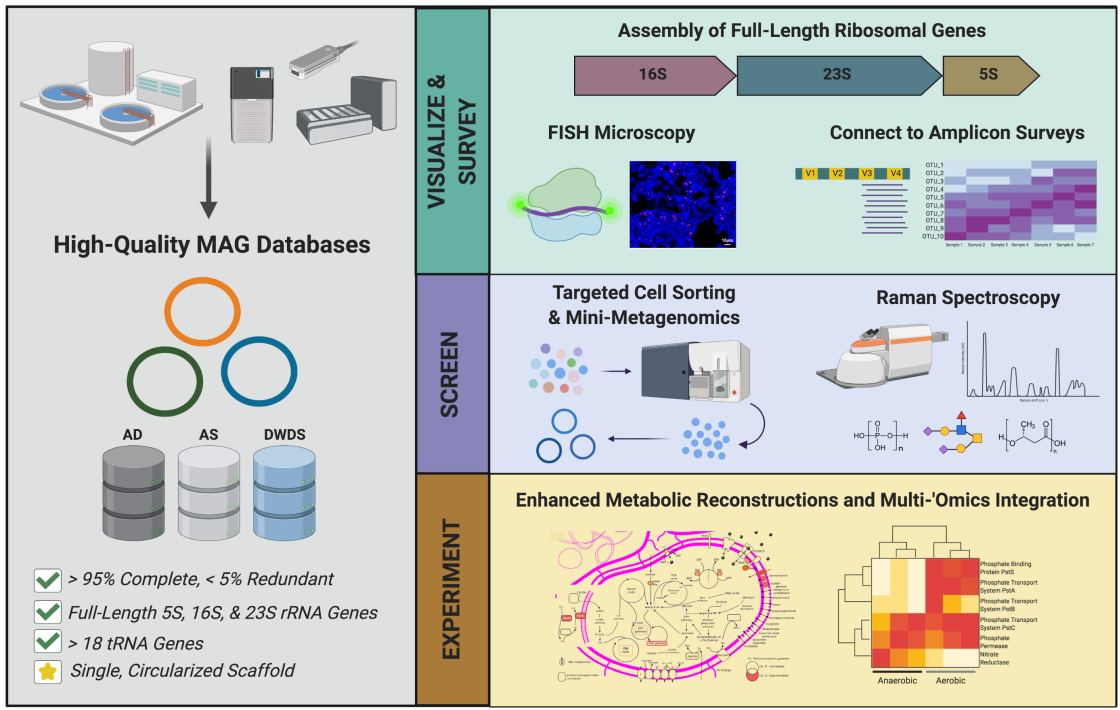

**Figure 6. Emerging technologies and future prospects for multi-'omics of engineered water microbiomes**: improvements and decreasing costs in long-read sequencing technologies paired with novel experimental technologies have paved the way for exciting discoveries in engineered water systems. With genome-resolved metagenomic sequencing surveys of AD, AS, and DWDS ecosystems, high-quality genome databases can be constructed for downstream experiments. The assembly of full-length ribosomal genes from MAGs will allow for enhanced FISH probe construction and comparisons to amplicon sequencing studies. These genomes can then be used for the basis for screening for specific activities by targeted cell sorting or Raman spectroscopy. Multi-'omics experiments can then be applied to understand the metabolism, regulation, and dynamics of key microbial guilds in these systems. We envision that these approaches will become commonplace and iterative of each other, such as demonstrated in Singleton et al. 2020. Created with BioRender.com.